\documentclass[twocol]{ametsocV6.1}
\usepackage{graphicx}
\usepackage{natbib}
\usepackage{hyperref}
\usepackage{natbib}
\usepackage{amsmath,mathtools}
\usepackage{adjustbox}
\hypersetup{
    colorlinks = true,
    urlcolor   = blue,
    citecolor  = black, 
}
\usepackage{color}
\usepackage{bm}
\usepackage{subcaption} 

\newcommand{\RomanNumeralCaps}[1]

\def\beq{\begin{equation}}
\def\eeq{\end{equation}}

\newcommand{\Ac}{\mathcal{A}}

\newcommand{\delsig}{\delta_{\sigma}}

\newcommand{\defn}{\stackrel{\text{def}}{=}}

\newcommand{\erfc}{\mathrm{erfc}}

\def\ee{\mathrm{e}}
\def\ii{\mathrm{i}}
\def\bx{{\bm{x}}}

\def\bk{{\bm{k}}}
\def\bU{\bm{U}}

\def\bq{\bm{q}}

\def\bP{\bm{P}}

\def\dd{\mathrm{d}}

\def\bHs{\bar {H}_s}
\def\bA{\bar \Ac}
\def\Hs{H_s}
\def\hs{h_s}

\def\hhs{\hat{h}_s}

\def\hlphi{\hat{{L}}_{\phi}}
\def\hlpsi{\hat{{L}}_{\psi}}
\def\eq{\bm{e}_{\bm{q}}}
\def\eqp{\bm{e}_{\bm{q}}^\perp}
\def\ellphi{\ell_\phi}
\def\ellpsi{\ell_\psi}

\def\qang{\varphi}

\def\spech{\hat{C}^{\hs}}

\renewcommand\Re{\mathop{\mathrm{Re}}}
\renewcommand\Im{\mathop{\mathrm{Im}}}
\newcommand\av[1]{\langle #1 \rangle}

\def\bcdot{\bm{\cdot}}

\newcommand\ch[1]{#1}

\definecolor{HW}{RGB}{137,0,225}

\newcommand{\KE}{K}

\title{The U2H map explains the effect of (sub)mesoscale \ch{currents} on significant wave height statistics} 

\authors{Han Wang\aff{1}, Ana B. Villas B\^{o}as\aff{2}, Jacques Vanneste\aff{3} and William R. Young\aff{4}}

\affiliation{
\aff{1} 
Institut f\"{u}r Meereskunde, Universit\"{a}t Hamburg, Hamburg 20146, Germany,
\aff{2}Department of Geophysics, Colorado School of Mines, Golden CO 80401, USA,
\aff{3}School of Mathematics and Maxwell Institute for Mathematical Sciences, University of Edinburgh, Edinburgh EH9 3FD, UK,
\aff{4} Scripps Institution of Oceanography, University of California at San Diego, La Jolla CA 92093-0213, USA
}

\abstract{Currents modulate the energy of surface gravity waves, leading to spatial inhomogeneities in significant wave height (SWH). 
Previous work indicates that the overall scale of the inhomogeneities is set by the scale of the currents, that the inhomogeneities are strongly anisotropic even for isotropic currents, and that the rotational and divergent components of the currents have sharply distinct effects. We explain these and other features of current-induced SWH inhomogeneities using the U2H map, a linear relation between SWH and currents deduced from wave-action conservation by making simplifying assumptions. We obtain a linear law relating the spectrum of SWH to the spectra of rotational and divergent kinetic energy of the current. This makes it possible to relate SWH statistics (such as variance and anisotropy) to the current statistics and wave properties including directional spreading.
}

\begin{document}

\maketitle

\section{Introduction} \label{sec:SWHturb}
Ocean surface gravity waves (SGWs) are primarily driven by wind but are  influenced by ocean currents, which modify their direction, frequency, and amplitude \citep{Phillips1977, Cavaleri2012}.  In recent years, recognition of the role of meso- and submesoscale turbulence in modulating SGWs has grown, driven by advances in numerical modeling \cite[e.g.,][]{rascle2014surface, Ardhuin2017, Romero2020, villas2020wave} and remote sensing observations \cite[e.g.,][]{rascle2016surface, Romero2017, Quilfen2019, lenain2021modulation}. While currents affect all moments of the SGW spectrum, recent studies have particularly focused on their influence on significant wave height (SWH), which is routinely observed by satellite altimeters and widely used in air--sea flux parametrizations \cite[e.g.,][]{taylor2001dependence, edson2013exchange}. Collectively, findings from this series of recent studies on wave--current interactions suggest the following:    

\begin{itemize}
\item[(i)]  rotational currents induce larger SWH anomalies than do divergent currents; 
\item[(ii)] the spatial scales of the SWH anomaly are determined by those of  the current, i.e.\ similar spectral slopes; 
\item[(iii)] the SWH response to divergent currents is local (largest where the current is strongest) but the response to rotational currents is non-local and  anisotropic.
\end{itemize}  
Despite  progress in our understanding of the relationship between currents and SWH anomalies, the exact mechanisms leading to points (i)--(iii) remain unclear.

\begin{figure*}
  \centering
  \begin{minipage}{\textwidth}
    \includegraphics[width=38pc]{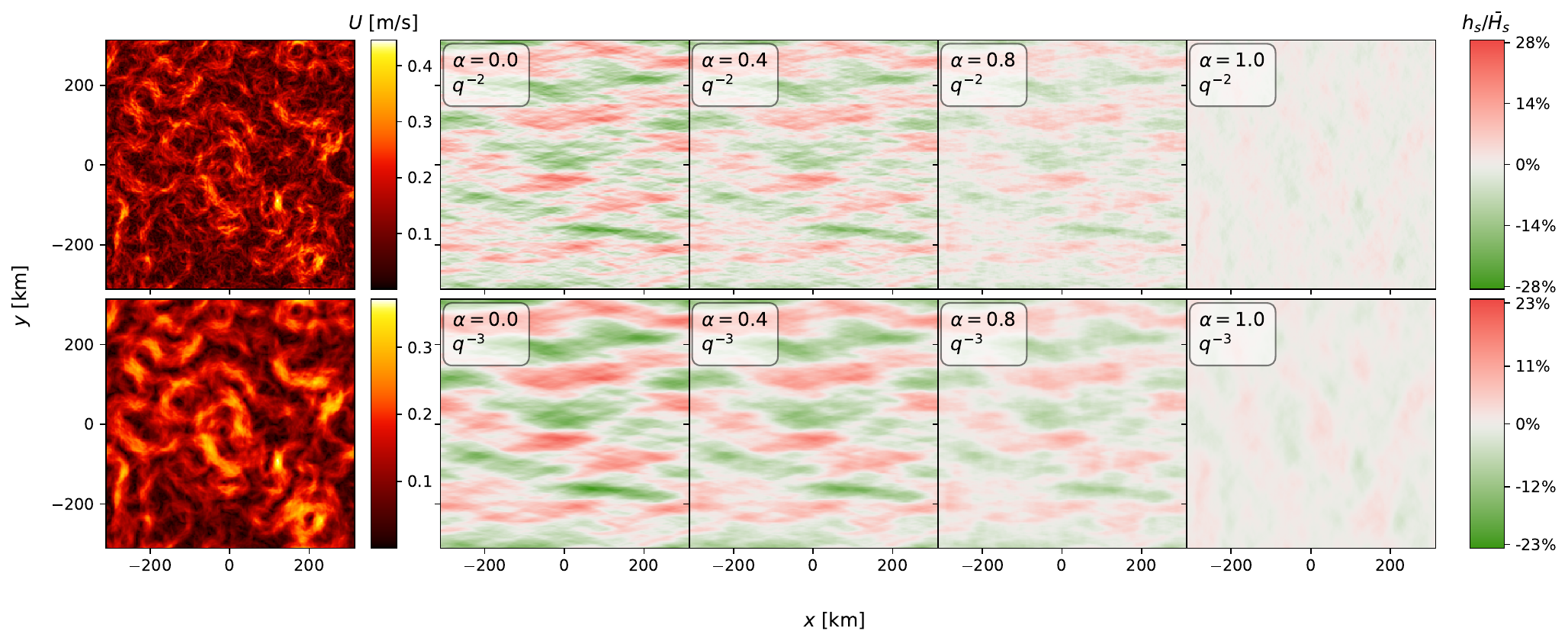}
    \caption{Impact of currents on SWH anomaly $\hs$.
    The current speed, $\sqrt{u^2+v^2}$, is shown in the first column. The KE spectrum of the current is isotropic and the isotropic spectra follow the power laws $q^{-2}$ (top row) and $q^{-3}$ (bottom row), where $q$ is the  wavenumber.   The relative SWH anomaly $\hs/\bHs$ is shown for four values of the parameter $\alpha$ defined in \eqref{betadef} controlling the ratio of rotational to divergent current \ch{KE} (columns two through five, with purely rotational current in column two and purely divergent current in column five). The waves have a wave action spectrum with directional-widthparameter $s=40$ in \eqref{LHCS} and the primary wave direction is along the $x$-axis. See section  \ref{sec:var} for details.
    }
    \label{fig:realsamplesingleslopes40}
  \end{minipage}

  \begin{minipage}{\textwidth}
    \includegraphics[width=38pc]{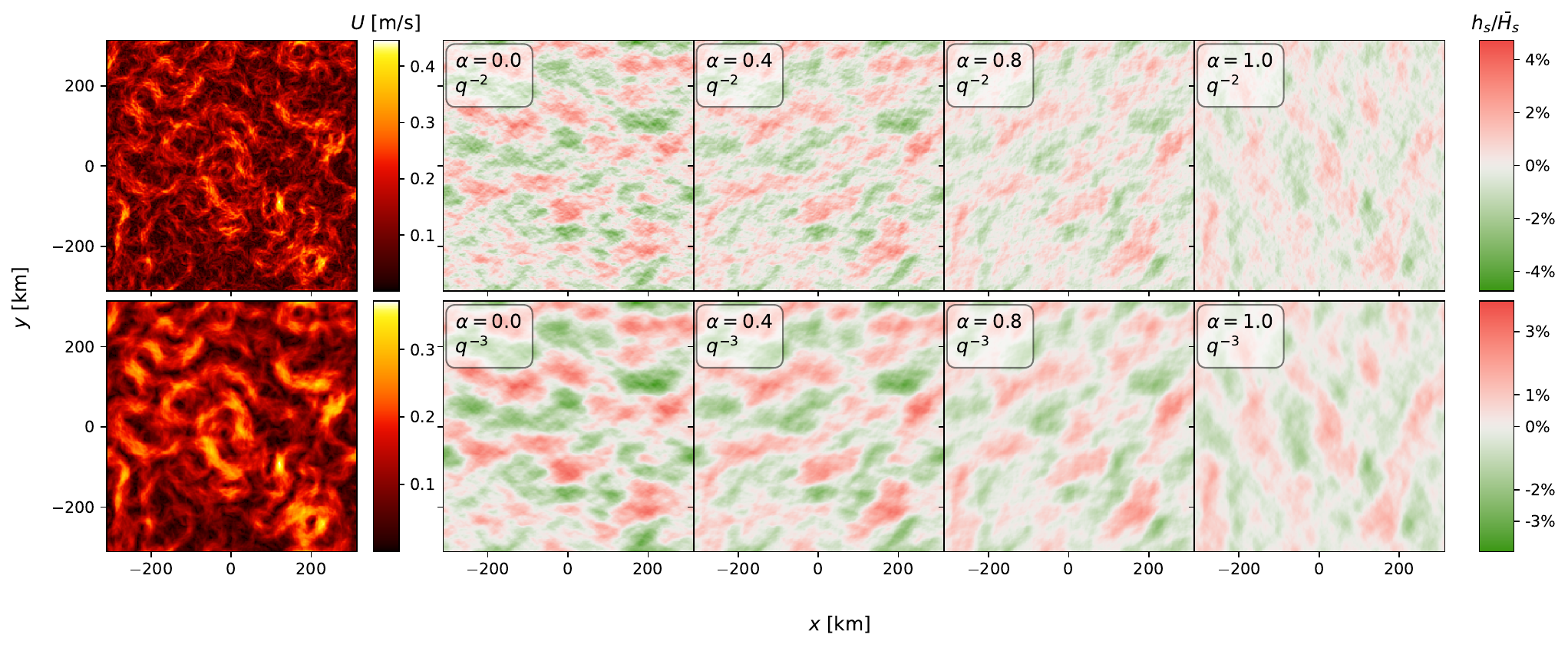}
    \caption{
    Same as figure \ref{fig:realsamplesingleslopes40} but  with directional-width parameter $s=1$ in \eqref{LHCS} corresponding to a \ch{spectrum with larger directional spreading}.
    Note the difference in color scale compared with figure  \ref{fig:realsamplesingleslopes40}.}
    \label{fig:realsamplesingleslopes1}
  \end{minipage}
\end{figure*}

Figure \ref{fig:realsamplesingleslopes40} shows the SWH anomaly, denoted $\hs(x,y)$, induced by synthetic random currents. The mean kinetic energy \ch{(hereafter ``KE")} of the currents is fixed, while the slope of the \ch{KE} spectrum and the ratio between the rotational and divergent energy of the current are varied. 
Features (i), (ii) and (iii) are evident: the magnitude of $\hs$ decreases as the proportion of rotational currents decreases, the spatial scales of $\hs$ are larger when the spatial scale of the current is larger, and the patterns of $\hs$ are stretched in the primary wave direction (chosen as the positive $x$-direction) when currents are  rotational. 

The main difference between figure \ref{fig:realsamplesingleslopes40} and figure 4 of \cite{villas2020wave} is the way in which SWH anomalies are computed: here we obtain $\hs$ using the U2H map developed in \citet[][hereafter W25]{wangU2H},
while in \cite{villas2020wave} $\hs$ is obtained by running WAVEWATCH III (WAVE height, WATer depth and Current Hindcasting, \citealt{tolman2009user} --  hereafter we refer to this model  as  WW3).
U2H is a linear map giving the SWH anomalies directly as a function of the current without resorting to solving the full wave-action conservation equation.
In W25, we compare the U2H map against WW3 in a few 
current configurations and establish a good match between the two methods.

While \cite{villas2020wave} focus their attention on \ch{wave spectra with small directional spreading, hereafter ``highly directional spectra"}, here we do not. We examine how the \ch{directional spreading} of the wave spectrum affects features (i), (ii) and (iii). This is illustrated in figure \ref{fig:realsamplesingleslopes1} which is similar to figure \ref{fig:realsamplesingleslopes40} but for a wave spectrum with \ch{large directional spreading}. For this \ch{spectrum}, features (i) and (ii) continue to hold, though less evidently for (i), but feature (iii) does not. Comparing figures \ref{fig:realsamplesingleslopes40} and \ref{fig:realsamplesingleslopes1} -- and noting the difference in the color maps --  shows that: 
\begin{itemize}
\item[(iv)] the SWH anomaly $\hs$ is larger for \ch{background waves with smaller directional spreading}.
\end{itemize}

Here we take advantage of
the simple, explicit form of U2H to explain and quantify features (i)--(iv). We also clarify the relation between  current and SWH spectra, refining our understanding of feature (iii), particularly where rotational and divergent currents have different spatial scales.

\section{U2H map: the recipe} \label{sec:U2Hrecipe}


The W25 recipe for the U2H map has two ingredients:
\begin{itemize}
\item the SGW action spectrum in the absence of currents, hereafter referred to as the background spectrum, $\bA (\bk)$, assumed independent of time and space;
\item the surface current velocity $\bU(\bx)$, assumed independent of time and with spatial scales much larger than SGW wavelengths.
\end{itemize}
The background significant wave height associated with $\bA (\bk)$ is
\beq \label{bHs2bA}
\bar \Hs  =  \left( \frac{16}{g} \int \! \bar \Ac(\bk) \sigma(k) \, \dd \bk \right)^{1/2},
\eeq
where $g$ is the gravitational acceleration, $k = | \bk |$, and 
\beq \label{dispersion}
\sigma(k)=\sqrt{gk}
\eeq 
is the intrinsic frequency of deep-water SGWs. $\bar \Hs$ is constant in space. 

The current $\bU(\bx)$ modulates the SGWs, leading to a spatially dependent SWH 
\beq
H_s(\bx) = \bHs + \hs(\bx),
\label{Hsexp}
\eeq 
where $\hs(\bx)$ is the current-induced SWH anomaly. 
The U2H map assumes that the typical current  speed $U$ 
is small compared to the group speed $c_g$ of the  waves:
 \beq \label{eps}
  U/c_g \ll 1.
 \eeq
U2H then provides an approximation to the relative SWH anomaly $\hs(\bx)/ \bar \Hs$ at order $O(U/c_g)$. 

The map is best  formulated using the Helmholtz decomposition of the current into purely divergent and purely rotational parts,
\beq
\bU (\bx) = \nabla \phi(\bx) + \nabla^\perp \psi(\bx), \label{Helmreal}
\eeq
where $\phi$ and $\psi$ are the velocity potential and streamfunction.  We introduce the Fourier transform
\beq
\hat \bU(\bq) = \int \!\!\! \bU(\bx) \ee^{- \ii \bq \bcdot \bx}\, \dd \bx 
\eeq
of the current velocity, with $\bq$ the Fourier vector (not to be confused with the SGW wavevector $\bk$). Eq.\ \eqref{Helmreal} becomes
\begin{align} 
\hat \bU (\bq) &= \ii \bq \hat \phi (\bq) + \ii \bq^\perp \hat \psi  (\bq) \notag \\
&= \hat U_\phi  (\bq) \eq +\hat  U_{\psi}  (\bq) \eqp, \label{HelmFT}
\end{align}
where $q=|\bq|$, $\bq^\perp$ is obtained from $\bq$ by rotation by $\pi/2$ about a vertical axis, and   $\eq =\bq/q$ and $\eqp = \bq^\perp/q$ are unit vectors. In terms of the polar representation
$\bq = q (\cos \qang, \sin \qang)$, $\bq^\perp = q(-\sin\varphi, \cos \varphi)$. 
The U2H map expressed in  Fourier space is then
\beq\label{U2HHelm}
\frac{\hhs(\bq)}{\bHs} = \hlphi(\qang) \, \widehat{U}_{\phi} (\bq)+ \hlpsi(\qang)  \, \widehat{U}_{\psi} (\bq).
\eeq
$ \hlphi(\qang)$ and $\hlpsi(\qang)$ are  determined by the background action spectrum and control the impact of the divergent and rotational parts of the current on $\hs$. (These functions are denoted as $\hat L_\parallel(\qang)$ and $\hat L_\perp(\qang)$ in W25, respectively.)
A crucial result is that $ \hlphi(\qang)$ and $\hlpsi(\qang)$
depend on the Fourier vector $\bq$ only through the polar angle $\qang$. They are given by 
\beq
 \hlphi(\qang) =-\frac{32}{g \bHs^2}\bm{P}  \bcdot \eq ,
\label{lparlper1}
\eeq 
and
\beq
\hlpsi(\qang) = \frac{16}{g \bHs^2} \left( \sum_{n=-\infty}^\infty n (-\ii)^{|n|} 2 \pi p_n \, \ee^{n \ii \qang} -  2 \bP \bcdot \eqp \right), 
\label{lparlper2}
\eeq
where 
\begin{align} \label{pndef}
2 \pi p_n &\defn  \int_0^{2\pi}\!\!\!\!\! \ee^{-\ii n \theta} \int_0^\infty \!\!\!\bA(k,\theta) k^2 \dd k \, \dd \theta. 
\end{align}
The wave momentum $\bP$ in \eqref{lparlper1} and \eqref{lparlper2} is
\beq
\bP = \begin{pmatrix} +\Re  2 \pi  p_1 \\  -\Im  2 \pi  p_1 \end{pmatrix}.
\label{Pb1}
\eeq

Equations \eqref{U2HHelm}--\eqref{Pb1} include all the formulas required to compute the U2H map from the background action spectrum $\bA(k,\theta)$ and current velocity $\bU(\bx)$. These formulas are linear in $\bU(\bx)$ and $\bA(k,\theta)$, and require minimal computational resources. The computational implementation is straightforward, as described and illustrated with a Jupyter notebook in W25.

In addition to \ch{assuming that $\bU$ and $\bA$ depend only on $\bx$ and $\bk$, that the spatial scale of $\bU$ is much larger than SGW wavelengths, and that $U/c_g$ is small,} the U2H map neglects \ch{the vertical shear of $\bU$, } forcing, dissipation and wave--wave interactions. \ch{These assumptions can be violated in realistic scenarios. 
The assumptions of large group speed and negligible forcing/dissipation often do not apply to short wind waves; as a result, the U2H map is applicable mainly to freely propagating surface waves with small steepness distant from their generation sites, e.g., swell.
The independence of background spectrum on time and space is also questionable for wave groups, and for rapidly evolving currents such as tides.
} 

In  examples shown here, we follow W25 and assume that the background spectrum takes a separable form
\beq \label{bAfD}
\bA (k,\theta)=f(k)D(\theta).
\eeq
We take $f(k)$ to be a truncated Gaussian distribution in intrinsic frequency concentrated  around a peak value $\sigma_m$ (see appendix A for details). We set $\sigma_m = 2 \pi / 10.3$ s$^{-1}$ for all the results of the paper except in section  \ref{sec:peakfreak} which considers the sensitivity to $\sigma_m$.

The angular dependence $D(\theta)$ follows the model of \citet[][hereafter LHCS]{LHCS1963}:
\beq
D(\theta) \propto \cos^{2s} \left( (\theta-\theta_p)/2\right),
\label{LHCS}
\eeq
where $\theta_p$ is the primary angle of wave propagation taken as $0$, and $s$ is the directional-width parameter, with large $s$ corresponding to a small directional spreading. 

\ch{The model of \eqref{bAfD}--\eqref{LHCS} is chosen for simplicity. We emphasize} that the U2H map can be employed with arbitrary background spectra, including \ch{spectra that are not separable in $k$ and $\theta$}, and that our qualitative conclusions do not depend on our specific choice of the LHCS model spectrum. \ch{Different forms of the background spectra affect U2H only through the integral in \eqref{pndef}. In the Jupyter Notebook published in W25, arbitrary wave spectra can be prescribed by users.} 
For highly directional spectra, in particular, the \ch{directional-width} parameter $s$ can be interpreted more generally \ch{for other spectral forms} as $s = 2 /\delta^2$, where $\delta \ll 1$ is the directional spreading \cite[e.g.][]{kuik1988method}. 

\begin{figure}
  \centering
  \includegraphics[width=19pc]{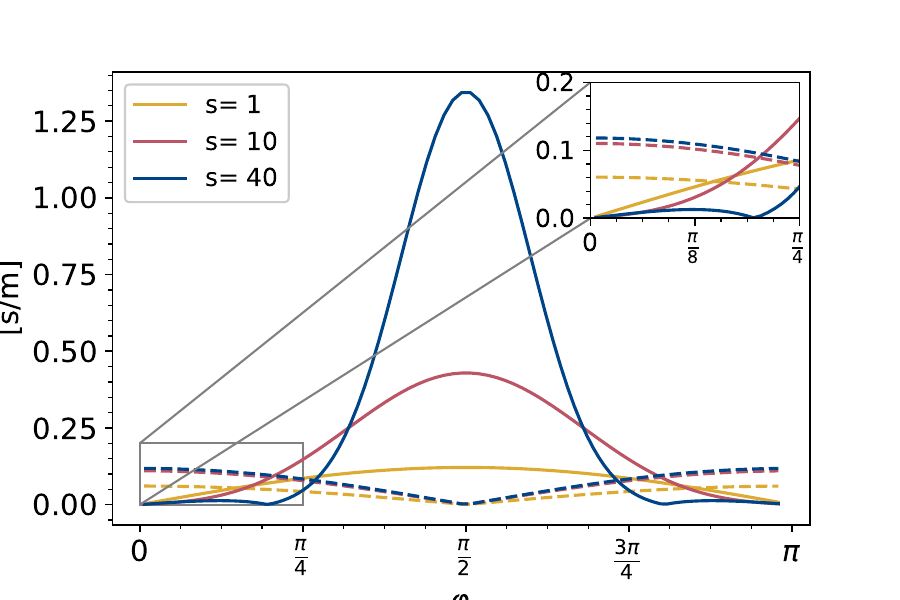}
    \caption{$|\hlphi|$ (dashed lines) and $|\hlpsi|$ (solid lines) computed from \eqref{pndef}--\eqref{Pb1} for the LHCS spectrum \eqref{LHCS} with $\theta_p = 0$ and for the directional-width parameter $s = 1, \, 10$ and $40$. A zoomed-in view for $0\leq\varphi<\pi/4$ is shown at the upper right corner.}
  \label{fig:Lparper}
\end{figure}

Figure \ref{fig:Lparper} shows $|\hlphi|$ and $|\hlpsi|$ for three values of $s$. Because $\hs$ and $\bU$ are real,
\beq\hlphi(-\qang)=- \hlphi^*(\qang) \quad \text{and} \quad \hlpsi(-\qang)=-\hlpsi^*(\qang).
\eeq
In figure \ref{fig:Lparper} we exploit this symmetry
to restrict the range of $\qang$ to $0<\varphi<\pi$.  For highly directional spectra, i.e.\ $s \gg 1$, $|\hlpsi|$ is localised around a maximum at $\varphi=\pi/2$ and is much larger than $|\hlphi|$. Asymptotic results reported in W25  explain these features by showing that $\hlpsi(\qang)$ is localised in $O(s^{-1/2})$ regions around $\varphi=\pm \pi/2$ where it scales like $s$. In contrast, $\hlphi(\qang)$ is $O(1)$ and $s$-independent for $s \gg 1$.


\section{Power spectral relations} \label{sec:spectralrelations}
W25 validate the U2H map \eqref{U2HHelm} by comparing its predictions to WW3 simulations for a few specific current configurations. In this paper we focus on the relationship between the statistics of SWH and the statistics of $\bU$ implied by \eqref{U2HHelm}. We assume that $\bU(\bx)$ is a homogeneous random field and define the spectra and cross-spectrum of the divergent and rotational parts of the current as
\begin{align} \label{CUphipsi1}
&\hat{C}^{U_{\phi}}(\bq) \defn  \langle |\hat U_{\phi}(\bq)|^2 \rangle
 = q^2 \langle |\hat \phi(\bq)|^2 \rangle, \\
 &\hat{C}^{U_{\psi}}(\bq) \defn  \langle |\hat U_{\psi}(\bq)|^2 \rangle = q^2 \langle |\hat \psi(\bq)|^2 \rangle , \label{CUphipsi2}
\end{align}
and
\beq
\hat{C}^{U_{\phi}U_{\psi}}(\bq)  \defn  \langle \hat U^*_{\phi}(\bq)\hat U_{\psi}(\bq) \rangle= q^2 \langle \hat \phi^*(\bq) \hat \psi(\bq) \rangle,
\label{spectra}
\eeq
where $\langle \cdot \rangle$ denotes ensemble average. 

From \eqref{U2HHelm} the SWH spectrum, 
defined as the power spectrum 
\beq
\spech(\bq) = \langle |\hhs(\bq)/\bHs|^2 \rangle
\eeq
of the relative change in SWH, is given by
\begin{align}
\label{U2Hspectra}
\spech(\bq) &= |\hlphi(\qang)|^2\hat{C}^{U_{\phi}}(\bq)+|\hlpsi(\qang)|^2 \hat{C}^{U_{\psi}}(\bq)  \notag\\&+
2\Re\left(\hlphi^*(\qang)\hlpsi(\qang) \hat C ^{U_{\phi}U_{\psi}}(\bq)  \right), 
\end{align}
where $\Re$ denotes the real part. 
Eq.\ \eqref{U2Hspectra} is central to this paper. It provides a simple relationship between the SWH spectrum and the spectra characterising the currents. 

We assume for simplicity that the currents are isotropic, with uncorrelated divergent and rotational parts. This reduces \eqref{U2Hspectra} to
\beq
\label{U2HspectraSimple}
\spech(\bq) = \frac{1}{\pi q} \left( |\hlphi(\qang)|^2 K_\phi(q)+|\hlpsi(\qang)|^2 K_\psi(q) \right),
\eeq
where $K_\phi(q)$ and $K_\psi(q)$ are the current divergent and rotational \ch{KE} isotropic spectra, with e.g. \begin{equation}
    K_\phi(q) = \tfrac{1}{2} \int \hat{C}^{U_{\phi}}(\bq) q \, \dd \qang = \pi q \hat{C}^{U_{\phi}}(q).
\end{equation}
The assumption of uncorrelatedness $\hat{C} ^{U_{\phi}U_{\psi}}= 0$ is not essential for many of our results, as discussed in appendix B.

While $\spech(\bq)$ is not isotropic because of the angular dependence of $\hlphi(\qang)$ and $\hlpsi(\qang)$, we can define the SWH isotropic spectrum
\beq
\label{defiso}
\spech(q) \defn \int \spech(\bq) q \, \dd \varphi,
\eeq
where we abuse notation in marking the difference between isotropic and two-dimensional spectra only by displaying the different independent variable, $q$ vs.\ $\bq$.
Integrating \eqref{U2HspectraSimple} over $\qang$  leads to 
\beq
\spech(q)  = 2 \left( \ellphi K_\phi(q) + \ellpsi K_\psi(q) \right)
\label{spectraiso} 
\eeq
where the coefficients $\ellphi$ and $\ellpsi$ are
\beq
(\ellphi\, , \ellpsi) = \frac{1}{2\pi} \int  \big(|\hlphi(\qang)|^2, \, |\hlpsi(\qang)|^2\big) \, \dd \varphi.
\label{eq:ll}
\eeq
We examine the consequences of \eqref{spectraiso} in the next two sections before considering the anisotropy of $\spech(\bq)$ in section \ref{sec:aniso}. 

\section{Variance} \label{sec:var}

We first study the variance of the SWH anomaly. Variance reflects the overall intensity of the anomaly, and helps understand features (i) and (iv) observed in section \ref{sec:SWHturb}.  

Integrating \eqref{spectraiso} with respect to $q$ and recalling from Parseval's identity that the variance of a homogeneous random field is the integral of its spectrum, we obtain a relation between the variance of the SWH and the variances of the divergent and rotational current speeds:
\beq \label{statvar}
\av{\hs^2/\bHs^2}= \ellphi \av{|U_{\phi}|^2}
+ \ellpsi \av{|U_{\psi}|^2}.
\eeq
Similarly, multiplying \eqref{spectraiso} by $q^2$ and integrating gives
\beq \label{statgrad}
\av{|\nabla\hs|^2/\bHs^2}  =
 \ellphi \av{D^2}  + \ellpsi \av{Z^2},
\eeq
where $D$ and $Z$ denote surface divergence and vorticity. 

Thus, the variance of SWH is a weighted sum of the variances of the speed of the divergent and rotational parts of the current. And the variance of the gradient of SWH is the weighed sum of the variances of the divergence and vorticity of the current. In both cases, the weights are $\ellphi$ and $\ellpsi$ in \eqref{eq:ll}.

\begin{figure}
  \centering
  \includegraphics[width=19pc]{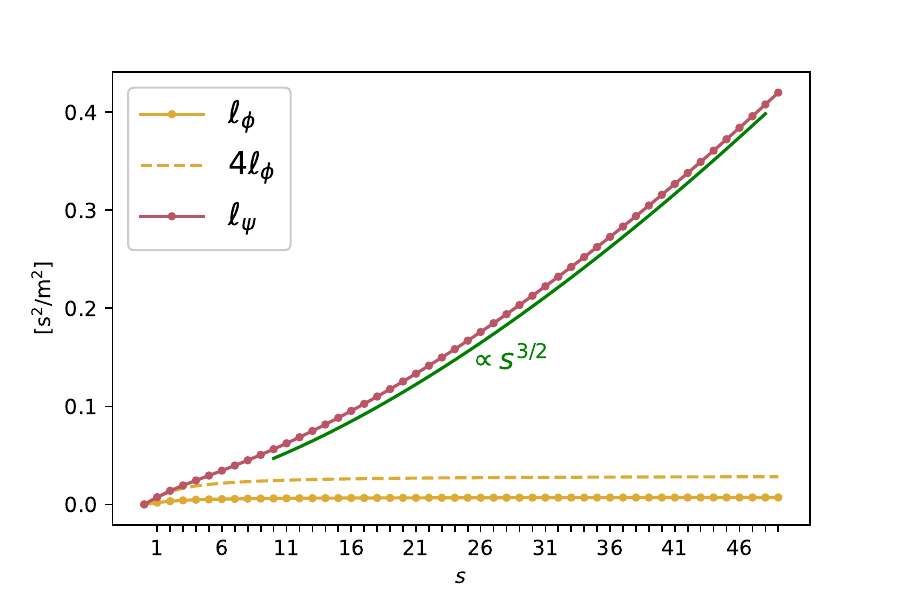}
    \caption{$\ellpsi$  and $\ellphi$ as functions of $s$, assuming the LHCS angular dependence. The green curve indicates the $s^{3/2}$ dependence of $\ellpsi$ that holds when $s\gg 1$.}
  \label{fig:Lint2}
\end{figure}

Figure \ref{fig:Lint2} shows $\ellpsi$ and $\ellphi$ for the LHCS model spectrum \eqref{LHCS} as functions of $s$. For $s=1$, $\ellpsi = 4\ellphi$. As $s$ increases, corresponding to \ch{smaller directional spreading}, $\ellpsi$ increases, as $s^{3/2}$ for $s \gg 1$ (this can be established from the large-$s$ form of $\hlpsi$ mentioned at the end of section \ref{sec:U2Hrecipe}), while $\ellphi$  saturates at a  much smaller value. These properties of $\ellpsi$ and $\ellphi$ explain features (i) and (iv) noted in section \ref{sec:SWHturb}, i.e.\ the dominance of rotational currents over divergent currents, and the increase in the magnitude of SWH anomalies \ch{for smaller directional spreading.} 

To quantity these effects, we express $\ellphi$ and $\ellpsi$ in terms of $p_n$ in \eqref{pndef} by substituting  \eqref{lparlper1} and \eqref{lparlper2} into \eqref{eq:ll} to find
\begin{align}
\ellphi &=\frac{1}{4}\left(\frac{32}{g \bHs^2}\right)^2 |\bP|^2\\
\ellpsi&=\left(\frac{32}{g \bHs^2}\right)^2 |\bP|^2 +\frac{1}{2}\left(\frac{32}{g \bHs^2} \right)^2  \sum_{|n|\geq2}n^2|p_n|^2. \label{ellsmodes}
\end{align}
Combining the two equations above 
\beq
\ellpsi = 4 \ellphi +\frac{1}{2}\left(\frac{32}{g \bHs^2} \right)^2  \sum_{|n|\geq 2}n^2|p_n|^2,
\label{eq:ellperellpar}
\eeq 
and therefore, as observed in Figure \ref{fig:Lint2},
\beq
\ellpsi \ge 4 \ellphi. \label{ellpsi4ellphi}
\eeq
 The large-$s$ behaviour of $\ellphi$ and $\ellpsi$ can be explained by the corresponding asymptotic form of 
$\hlphi(\qang)$ and $\hlpsi(\qang)$ (see W25).

\begin{figure}
  \centering
  \includegraphics[width=19pc]{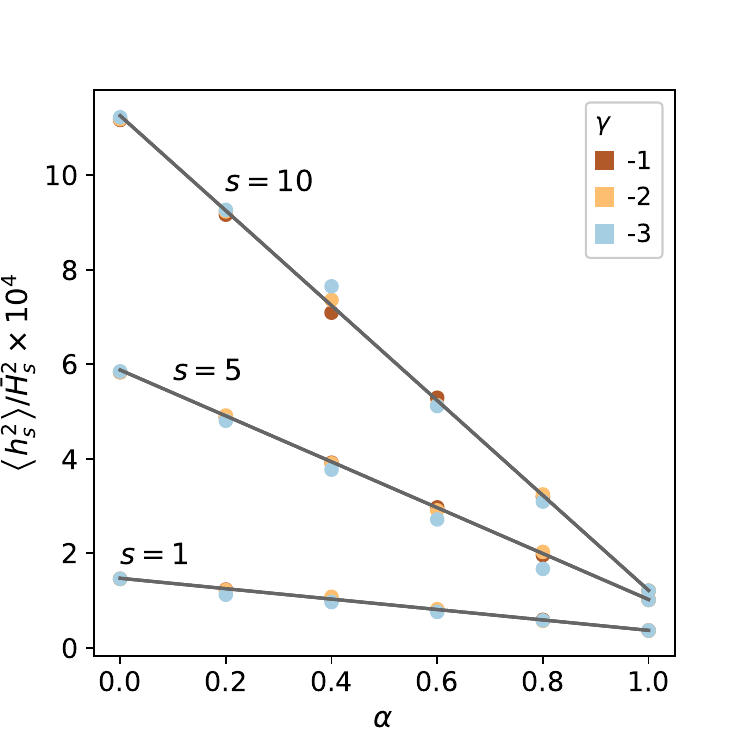}
    \caption{Variance $\left<\hs^2\right>/\bHs^2$ of the SWH anomaly as a function of the parameter $\alpha$ controlling the ratio of divergent to rotational \ch{KE} of the current. Three values of the spectral slope $\gamma=-1,\, -2$ and $-3$ are considered. Estimates from random samples (dots)   are compared with the prediction \eqref{statvar} (lines).}
  \label{fig:scatter_hsvar}
\end{figure}

We illustrate \eqref{statvar} in Figure \ref{fig:scatter_hsvar}.
This shows the SWH variance $\av{|\hs|^2/\bHs^2}$ computed for an ensemble of synthetic, randomly drawn currents. For all currents, the isotropic \ch{KE} spectra $\KE_\phi(q)$ and $\KE_\psi(q)$ have the same shape, corresponding to a power law $q^{\gamma}$  over a finite range of $q$ with isotropic spectral slope $\gamma = -1, \, -2$ and  $-3$ (see  Appendix C for details). The magnitudes of $\KE_\phi$ and $\KE_\psi$
are varied in such a way that the total kinetic energy $\KE=\KE_\phi + \KE_\psi$ is a constant ($0.01$ m$^2/$s$^2$ per kilogram). Figure \ref{fig:scatter_hsvar} shows $\av{|\hs|^2/\bHs^2}$ as a function of the parameter $0 \le \alpha \le 1$ such that
\beq
\KE_\phi(q)  = \alpha \KE(q) \quad 
\textrm{and} \quad 
\KE_\psi(q)  = (1-\alpha) \KE(q), \label{betadef}
\eeq
and for three values of $s$. As expected from \eqref{statvar}, the SWH variance $\av{|\hs|^2/\bHs^2}$  is independent of the spectral slope $\gamma$, linear in $\alpha$, and increasing with $s$ (since combining \eqref{statvar} with \eqref{betadef} shows the SWH variance is proportional to $\alpha \ellphi + (1-\alpha) \ellpsi$). 
We have checked that the SWH gradient variance behaves as indicated by \eqref{statgrad}.

\section{Isotropic spectrum}\label{sec:isospec}

We now examine the implications of formula \eqref{spectraiso} for the SWH isotropic spectrum $\spech(q)$. According to \eqref{spectraiso}, $\spech(q)$ is the weighted sum of the divergent and rotational \ch{KE} spectra $K_\phi(q)$ and $K_\psi(q)$, with $q$-independent weights $\ellphi$ and $\ellpsi$. This explains feature (ii) of section  \ref{sec:SWHturb}:  the length scales of $\hs$ are set by those  of the current. 

When the divergent and rotational spectra share an identical spectral slope, as in \cite{villas2020wave} and in figures \ref{fig:realsamplesingleslopes40} and \ref{fig:realsamplesingleslopes1}, it  follows from \eqref{spectraiso} that $\spech(q)$ has the same spectral slope. 
In typical oceanic flows, however, the rotational spectrum $\KE_\psi(q)$ is steep, dominating the total KE at large scales, and the divergent spectrum $\KE_\phi$ is shallower, dominating the total KE at small scales \citep{Buhler2014,RochaDrake2016, balwada2016scale,Chereskin2019}.
For such flows \eqref{spectraiso} predicts that $\spech(q)$  is parallel to $\KE_\psi(q)$ at large scales and to $\KE_\phi(q)$ at small scales. Because $\ellpsi > 4\ellphi$, the \ch{spectral slope of $\spech(q)$ is steeper than or as steep as $K(q)$, and the }transition scales between steep and shallow spectra arises at smaller scales for  $\spech(q)$ than for $K(q)$. The shift towards small scales is particularly pronounced for waves with \ch{small directional spreading} ($s \gg1$) for which $\ellpsi \gg \ellphi$.

\begin{figure*}[ht]
  \centering
  \includegraphics[width=19pc]{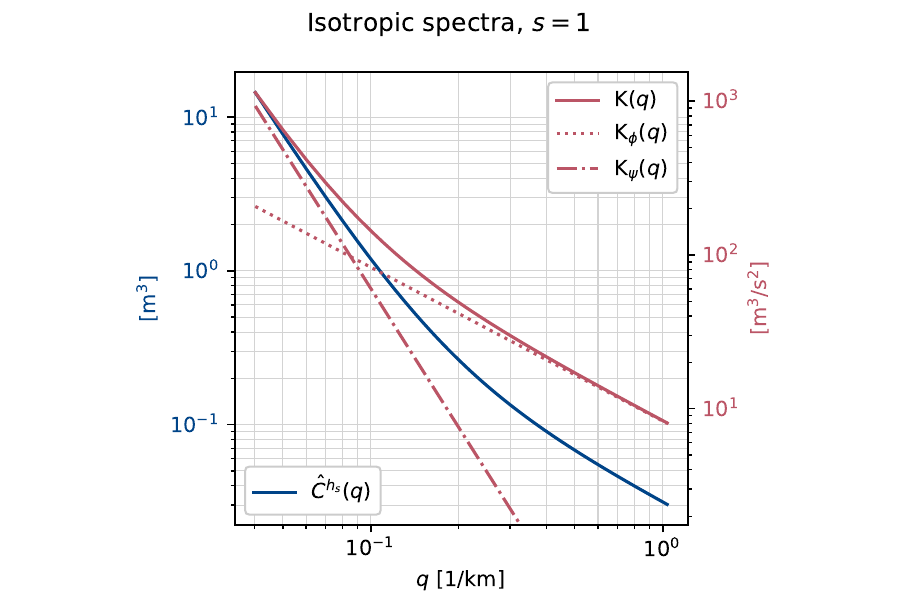}
    \includegraphics[width=19pc]{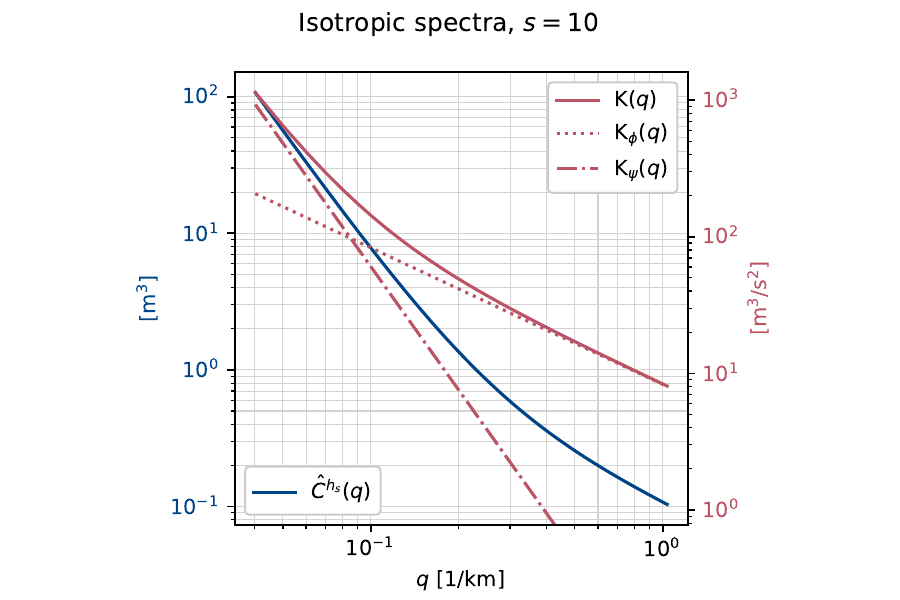}
    \caption{Isotropic spectrum $\spech(q)$ and  KE spectra (total, divergent and rotational KE) for the examples  in section \ref{sec:isospec}. The wave directional-width parameter $s$ is $1$ (left panel) and $10$ (right panel). The left vertical axis is for $\spech(q)$ and the right vertical axis is for the KE spectra. Spectra outside $(200 \text{km}) < q < 2\pi/(6 \text{km})$ are not shown.} 
  \label{fig:isospec}
\end{figure*}


In Figure \ref{fig:isospec} we compare the SWH spectrum $\spech(q)$ to  the divergent, rotational and total \ch{KE} spectra of the currents for $s=1$ (left panel) and $s=10$ (right panel). The energy spectra $\KE_\phi(q)$ and $\KE_\psi(q)$ are taken as the power laws $q^{-1}$ and $q^{-3}$, respectively, with amplitudes such that the wavenumber where $\KE_\psi(q)=\KE_\phi(q)$ is  $q=2\pi/(100$ km). 
(These parameters are chosen for pedagogical purpose and are not intended to agree with realistic currents.) The background wave spectrum is as in section  \ref{sec:var}. The spectrum $\spech(q)$ is obtained by computing $\hs(\bx)/\bHs$  for 10 random current realizations using the U2H map and averaging. This spectrum is as expected, making a transition from steep to shallow at a \ch{higher wavenumber (smaller spatial scales)} than the transition in the total \ch{KE} spectrum. In the case \ch{with small directional spreading} $s=10$, the transition is shifted 
\ch{to even higher wavenumbers.}
\ch{The spectral slope of $\spech(q)$ is overall steeper than $\KE(q)$, with the difference more pronounced for smaller directional spreading. The steeper spectral slopes of $\spech(q)$ are also visible in  }
results reported in \cite{Ardhuin2017}. 

\begin{figure*}[ht]
  \centering
  \includegraphics[width=38pc]{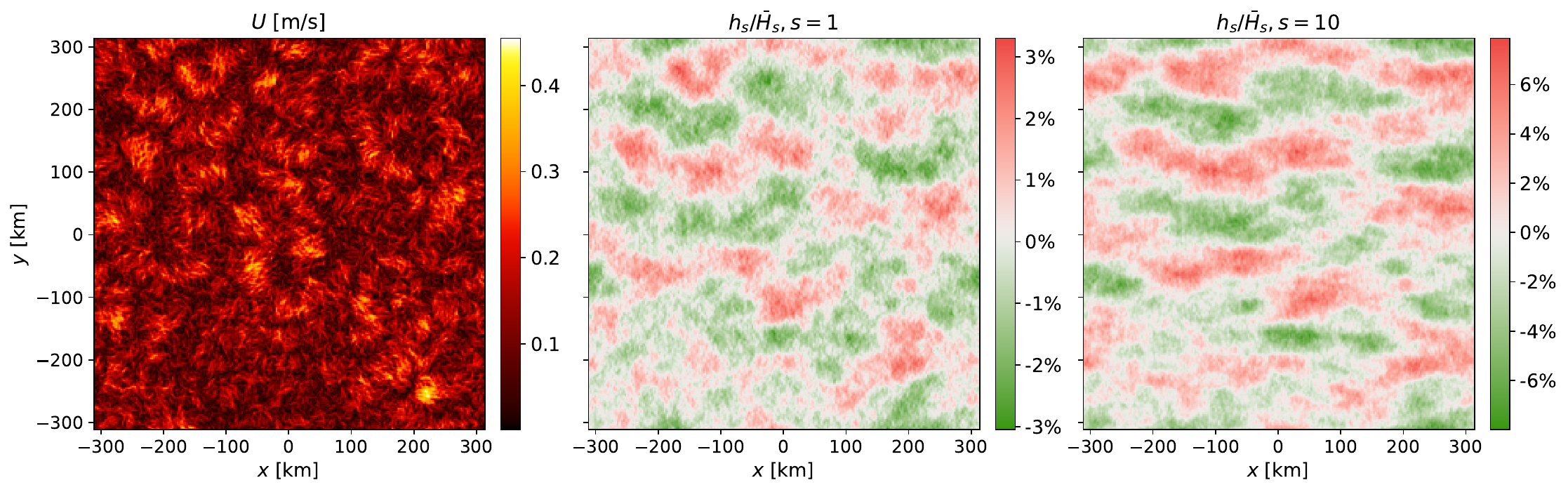}
    \caption{SWH anomaly $\hs/\bHs$ (middle and right panels) induced by  a current sample described in section \ref{sec:isospec}. The left panel shows the current speed. The background waves follow the LHCS model described in appendix A, with directional-width parameter $s$ taken to be $1$ (middle panel) or $10$ (right panel). }
  \label{fig:realspacesamples}
\end{figure*}

For currents such as those of Figure \ref{fig:isospec}, dominated by rotation at large scales and by divergence at small scales, the \ch{steeper spectral slope }
of $\spech(q)$ implies that the overall scale of the SWH anomaly $\hs$ is larger than the scale of the current, markedly so for \ch{waves with small directional spreading}. This is confirmed by 
Figure \ref{fig:realspacesamples} which shows $\hs$ in physical space for a single realization of the currents with the same parameters as for Figure \ref{fig:isospec}. The anisotropy of $\hs$ is striking and demands quantification.

\section{Anisotropy} \label{sec:aniso}

We first examine the anisotropy of the SWH anomaly by contrasting the one-dimensional spectra
\beq\label{defalong}
\spech(q_1)= \frac{1}{2\pi }\int \spech(q_1,q_2)  \, \dd q_2 
\eeq
and
\beq\label{defacross}
\spech(q_2)=\frac{1}{2\pi }\int \spech(q_1,q_2)  \, \dd q_1
\eeq
corresponding to observations of $\hs$ along or across the primary wave direction.  
\begin{figure*}[ht]
  \includegraphics[width=19pc]{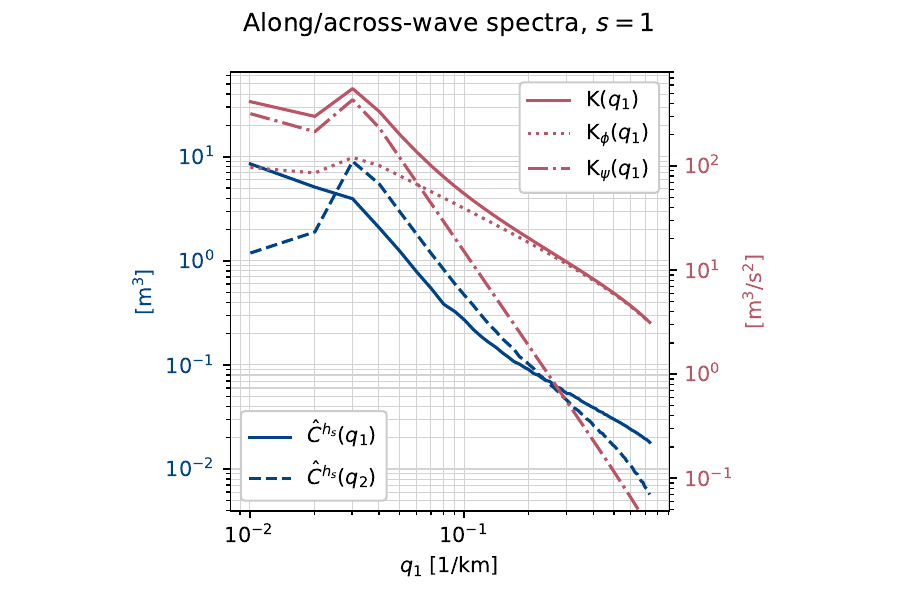}
  \includegraphics[width=19pc]{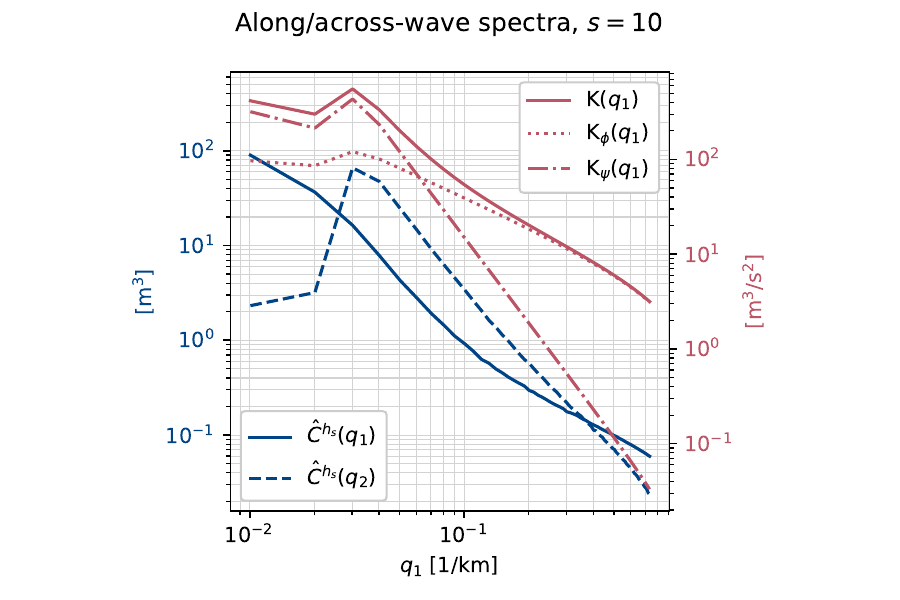}
    \caption{Along/across-wave SWH spectra $\spech(q_1)$ and $\spech(q_2)$, and total, divergent and rotation KE spectra as functions of wavenumbers $q_1$ or $q_2$. The left vertical axis is for $\spech(q_1)$ and the right vertical axis is for the KE spectra. The
wave directional-width parameter $s$ is 1 (left panel) and 10 (right panel). Spectra outside $(200 \text{km}) < q < 2\pi/(6 \text{km})$ are not shown.}
  \label{fig:hsKE_bislope}
\end{figure*}

Figure \ref{fig:hsKE_bislope} shows these spectra together with analogous KE spectra for the current and wave parameters of Figure \ref{fig:isospec} and for $s=1$ and $10$. 
In the along-wave direction,  $\spech(q_1)$ peaks
at the lowest wavenumbers regardless of the spectral slope of the KE spectra. The slope of $\spech(q_1)$ for small $q_1$ increases with $s$, corresponding to increasingly large scales dominating the along-wave SWH signal. 

In the across-wave direction, $\spech(q_2)$ peaks at scales close to the KE peak, but has a spectral slope closer to the rotational KE (steeper than the total KE for realistic oceanic currents) and thus  leading to  SWH scales larger than the current scales.
These properties of $\spech(q_1)$ and $\spech(q_2)$ reflect the streaky nature of the SWH anomaly seen in Figures \ref{fig:realsamplesingleslopes40}, \ref{fig:realsamplesingleslopes1} and \ref{fig:realspacesamples}.
They can be fully understood by combining the closed-form expression \eqref{defalong}--\eqref{defacross} with the explicit expression
\eqref{U2HspectraSimple} for $\spech(\bq)$ and the form of $\hlphi(\qang)$ and $\hlpsi(\qang)$ shown in Figure \ref{fig:Lint2}. 
In the limit of large $s$, the asymptotic behaviors seen in  Figure \ref{fig:Lparper} (discussed in section \ref{sec:U2Hrecipe}) amplify these properties, leading to more streaky SWH anomalies.

\begin{figure}[ht]
  \centering
  \includegraphics[width=19pc]{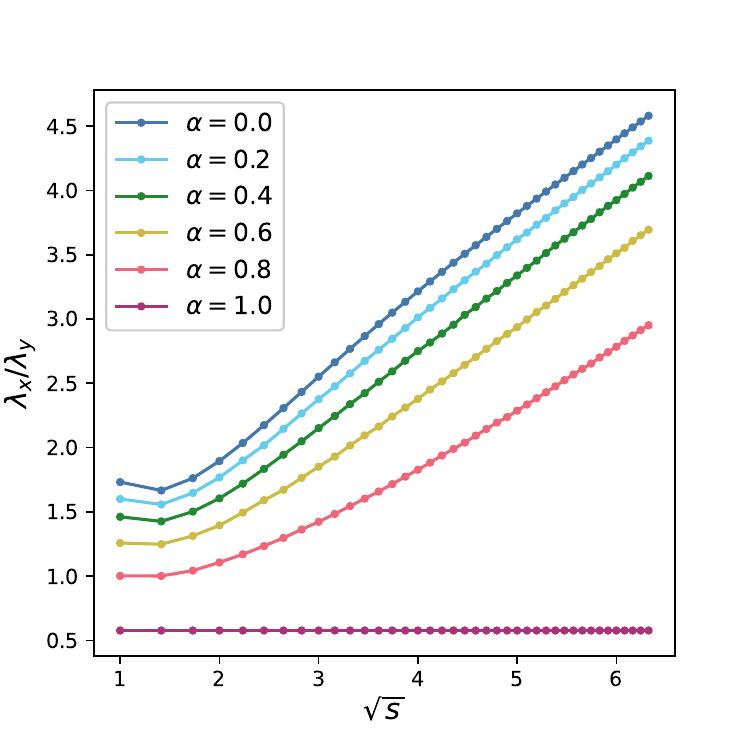}
    \caption{Aspect ratio $\lambda_x/\lambda_y$ of the SWH anomaly as a function of the directional-width parameter $s$ and for different values of the parameters $\alpha$ controlling the ratio of rotational to divergent \ch{KE} of the current.}
  \label{fig:aspecratio}
\end{figure}

A simple measure of the anisotropy of $\hs$ is the aspect ratio $\lambda_x/\lambda_y$ of typical lengthscales in the along- and across-wave directions. This ratio can be defined by 
\beq
\left(\frac{\lambda_x}{\lambda_y}\right)^2=\frac{\av{(\partial_y \hs)^2}}{\av{(\partial_x \hs)^2}} = \frac{
\iint q_2^2 \spech(\bq)\,\dd \bq} {
\iint q_1^2 \spech(\bq)\,\dd \bq}, \label{aspecratiosq}
\eeq
In figure \ref{fig:aspecratio}, we show $\lambda_x/\lambda_y$ as a function of the \ch{directional-width parameter} $s$ and for different values of the parameter $\alpha$ characterizing the relative importance of the divergent and rotational parts of the current. The other parameters are a described in section \ref{sec:var}, with the spectral slope fixed as  $\gamma  =  -3$. Except for purely divergent currents ($\alpha = 1$), $\lambda_x/\lambda_y$ is greater than $1$, reflecting elongation of the SWH patterns in the along-wave direction, and increasing with $s$ for $s \gtrsim 1.5$. 

For $s \gg 1$, $\lambda_x/\lambda_y \propto \sqrt{s}$, as is established by noting that
\beq
\left(\frac{\lambda_x}{\lambda_y}\right)^2 \propto \frac{\int \sin^2{\varphi}|\hlpsi(\qang)|^2 \,\dd \varphi}{\int \cos^2{\varphi}|\hlpsi(\qang)|^2 \,\dd \varphi} \propto s,
\label{scaleratio}
\eeq
since $\hlpsi \gg \hlphi$ and $\hlpsi$ is localized within an $O(s^{-1/2})$ boundary layer around $\qang = \pi/2$. This provides a scaling for the elongation of the streaks in $\hs$ for \ch{waves with small directional spreading}.

For purely divergent currents ($\alpha=1$), the aspect ratio is a constant (because $\hlphi \propto \cos \varphi)$, computed to be $\lambda_x/\lambda_y=0.58 < 1$. This is why, in the rightmost $\alpha =1.0$  panels of figures \ref{fig:realsamplesingleslopes40} and \ref{fig:realsamplesingleslopes1} (with $\alpha = 1$),  $\hs$ is somewhat  elongated in the across-wave $y$-direction.


\section{Sensitivity to mean frequency} \label{sec:peakfreak}

In preceding  numerical examples we  use mean radian frequency $\sigma_m = 0.61$ s$^{-1}$, corresponding to the mean period $T_m = 10.3$~s.  The background action spectrum $ \bA(k,\theta)$ depends on $\sigma_m$ (see appendix C) and  the U2H functions $\hlphi(\qang)$ and $\hlpsi(\qang)$  inherit this dependence   through the integral
\begin{equation}
\int_0^\infty \!\!\! \bA(k,\theta) k^2 \dd k
\end{equation}
in \eqref{pndef}.
In this section, we examine the sensitivity of $\ellphi$ and $\ellpsi$ to $\sigma_m$.

\begin{figure*}
  \centering
  \includegraphics[width=38pc]{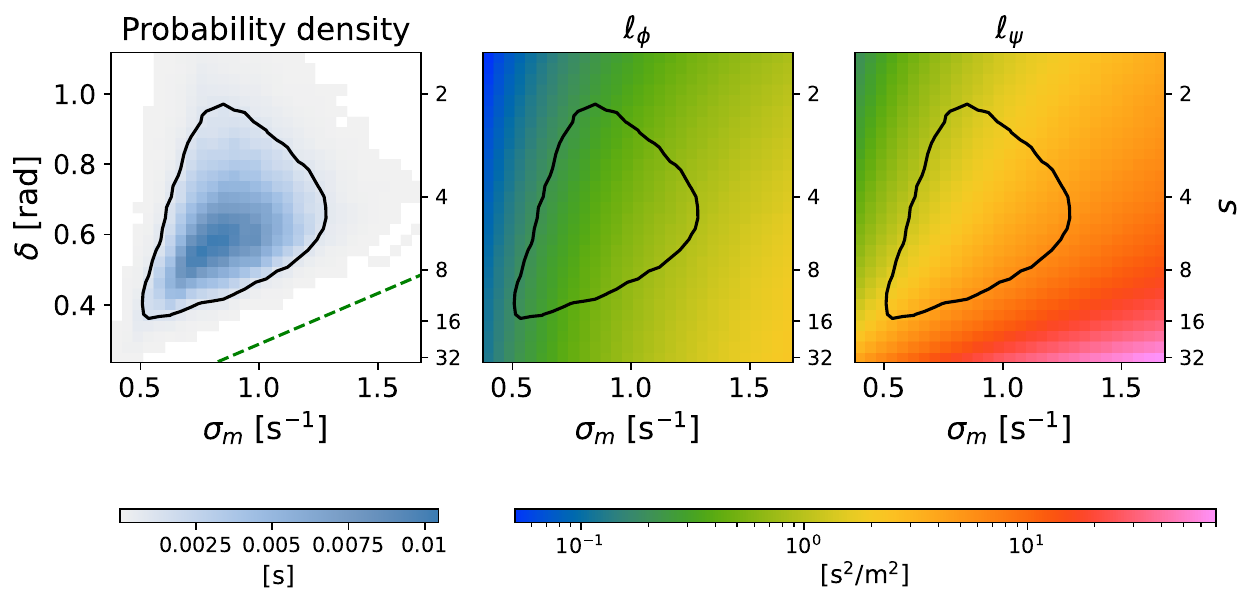}
    \caption{Joint probability density of wave frequency $\sigma_m$ and directional spreading $\delta$ from Ocean Station Papa observations (left panel) and U2H functions $\ellphi$ (middle panel), and $\ellpsi$ (right panel) as functions of  $(\sigma_m,\delta)$, based on the LHCS model and the relation $\delta =  (2/s)^{1/2}$. The probability density is constructed from $30\times30$ bins equally spaced in $\delta$ and $\sigma_m$ with the color indicating the fraction of samples in each bin.  The solid contour encloses bins with probability densities greater than $0.001$ s;  these bins enclose $92\%$ of the data. Some values of $s$ corresponding to $\delta$ are marked on the right vertical axes. The green dashed line in the left panel indicates the boundary of the linear regime (see \eqref{linearity} in Discussion).}
  \label{fig:s_fp_contour} 
\end{figure*}

In ocean-relevant situations, the dependence of $\ellphi$ and $\ellpsi$ on $\sigma_m$ is comparable to the dependence on the directional spreading parameter $s$. This is inferred from figure \ref{fig:s_fp_contour} which shows $\ellphi$ and $\ellpsi$ as functions of both $\sigma_m$ and $s$ for wave spectra with fixed total energy, equivalently fixed $\bHs$. For oceanographic context, the leftmost panel of figure \ref{fig:s_fp_contour} shows an estimated probability density of $\sigma_m$ and $\delta$ from 13 years (2010--2023) of wave buoy observations from Ocean Station Papa in the North Pacific Ocean. We choose Ocean Station Papa as a reference location due to its exposure to a broad range of wave conditions and its long history of meteorological and oceanographic observations \cite[e.g.,][and references therein]{large1981open, thomson2013waves}. Directional wave buoy observations routinely provide the first four Fourier coefficients of the directional energy distribution, from which the mean frequency $\sigma_m$ and mean directional spreading $\delta$ can be computed 
\citep{kuik1988method}. There are $195\,580$ samples included. To aid visualization, extreme observations with $\delta<0.183$, $\delta>1.118$, or $\sigma_m> 1.7$~s$^{-1}$, are deleted  ($0.5\%$ of the data is discarded). The probability density plot indicates the range of typical ocean values of $\delta$ and $\sigma_m$. As expected, short waves (large $\sigma_m$) \ch{have larger directional spreading} (larger $\delta$) than long waves. 

The second and third panels of figure \ref{fig:s_fp_contour} show $\ellphi$ and $\ellpsi$ computed by assuming the LHCS spectra for background waves, with directional spreading parameter $s=2/\delta^2$. In summing the Fourier series \eqref{lparlper2}, we cut $n$ at $\pm50$, after verifying that  $n=100$ does not significantly change the results. 
For all relevant parameter values, $\ellphi \ll \ellpsi$.  $\ellphi$ and $\ellpsi$ in \eqref{eq:ll} increase with $\sigma_m$ or, equivalently, with mean wavenumber $k_m = \sigma_m^2/g$, i.e.\ short high-frequency waves are more affected by currents and produce larger SWH anomalies than long low-frequency waves. 
The changes in $\ellpsi$ produced by varying $\sigma_m$ over the data range are comparable to those resulting from variations in $s$. For $s \gg 1$, in particular,  we can show that $\ellpsi \propto s^{3/2} \sigma_m^2$. The $s^{3/2}$ was noted in section \ref{sec:var}; the $\sigma_m^2$ stems from the approximation 
\begin{align}
\int_0^\infty \bA(k,\theta) k^2 \dd k 
&\sim k_m^{1/2} \int_0^\infty \bA(k,\theta) k^{3/2} \dd k \nonumber \\
&\sim k_m^{1/2} \bHs^2 \propto \sigma_m \bHs^2  . 
\end{align}
and the quadratic dependence of $\ellpsi$ on the left-hand side (see \eqref{lparlper1}--\eqref{lparlper2} and \eqref{eq:ll}).

Earlier discussion emphasized the sensitivity of $\ellpsi$ to changes in $s$. Here we emphasize that $\ellpsi$ is equally sensitive to changes in mean frequency $\sigma_m$.  The same conclusion applies to the function $\hlpsi$ and hence to the SWH anomaly spectra.

\section{Discussion} \label{sec:discussion}


Key statistical features of current-induced SWH are explained by the U2H map which provides a linear relation between the anomalies of SWH and currents. U2H is deduced from the wave-action conservation equation by making a number of assumptions, chiefly the assumption \eqref{eps} of current speeds small compared with wave group speeds. The physical effects captured by U2H identify the main mechanisms by which currents induce SWH anomalies. 

In U2H, advection of the action spectrum by currents is of secondary importance  and is entirely neglected on the basis of $U/c_g \ll 1$. (See \citet{Ardhuin2017} for simulation results with and without advection.)
Instead, currents affect the distribution of wave action, and hence of SWH,  by inducing changes in both the direction and wavelength of SGWs (through refraction and the ``concertina effect'', respectively).  


In U2H, refraction and concertina effects superpose linearly. It is straightforward to trace their contribution through the derivation of W25: refraction is associated with directional variations $\partial_\theta \bA$ of the background action and concertina effect with wavenumber variations $\partial_k \bA$. Refraction can then be shown to contribute to the series in expression \eqref{lparlper2} for $\hlpsi$. 
The remaining term in $\hlpsi$ and the entirety of $\hlphi$ are proportional to $\bm{P}$, which can be attributed to a combination of  refraction and concertina effect, with respective weights $-1/2$ and $3/2$. For highly directional \ch{spectra}, $|\hlpsi| \gg |\hlphi|$ and $\hlpsi$ is dominated by the series in \eqref{lparlper2}, so we conclude that refraction, driven by vorticity, is the primary mechanism of current-induced SWH anomalies. This is consistent with the classical result showing that under asssumption \eqref{eps} ray curvature is determined by vorticity \citep{LLfluid,Kenyon1971,Dysthe2001,Gallet2014}.

While this paper focuses on statistical predictions, we emphasize that U2H provides
an approximation to the SWH anomaly induced by any specific current, as figures \ref{fig:realsamplesingleslopes40} and  \ref{fig:realsamplesingleslopes1} illustrate. We also emphasize that the statistical assumptions we make (isotropy of the currents and, in some cases, independence of divergent and rotational currents) are not required by U2H. Rather, we make them to obtain simple, explicit predictions that explain generic features of wave--current interactions. The regime of \ch{waves with small directional spreading} is particularly robust to details of the current spectrum because the SWH anomaly is then controlled by the spectrum of the rotational current $\hat{C}^{U_{\psi}}(\bq)$ in a narrow wedge $\qang \approx \pm \pi/2$ in $\bq$-space and insensitive to both $\hat{C}^{U_{\phi}}(\bq)$ and $\hat{C}^{U_\phi U_{\psi}}(\bq)$.

\ch{The framework of U2H can be generalized to map currents to current-induced anomalies of other wave quantities computable from wave action spectra, including directional spreading, mean periods, and the Stokes drift.}

\ch{U2H depends on a linear approximation and requires that  $\hs/ \bar \Hs = O(U/c_g)\ll1$. To ensure that $\hs$ is small relative to $\bar \Hs$, some restrictions are needed. In particular, small directional spreading leads to strong SWH anomalies, calling for an upper bound on $s$.  }
W25 discuss this and show that condition \eqref{eps} needs tightening to
\begin{equation}  \label{linearity}
U s^{1/2} / c_g \ll 1.
\end{equation}
We show the curve $U s^{1/2} / c_g =  1$ or equivalently, in terms of the directional spreading $\delta$ and mean frequency $\sigma_m$, $2\sqrt{2}U\sigma_m=g\delta$
for $U = 1$ m s$^{-1}$ in the probability density plot in figure \ref{fig:s_fp_contour} (left panel). All data points are well above this green dashed line, indicating that the inequality \eqref{linearity} is satisfied. This suggests that \eqref{linearity} is only violated in rare circumstances, e.g.\ when current speeds are well in excess of $1$ m s$^{-1}$. 


\ch{U2H is not applicable to length scales larger than $O(1000)$ km. Over large propagation distances, the cumulative scattering by currents leads to substantial broadening of the SGW spectrum, challenging the assumption that $\bA$ is independent of space. }
\citet{Smit2019} and \citet{VillasBoas2020b}  show that this broadening amounts to a directional diffusion of the action, ultimately leading to isotropic spectra (diffusion in frequency is weak as it depends on  the time-dependence of the currents, see \citet{cox2023inertia}). The comparison between U2H and WW3 simulations in \ch{\citet{wangU2H}} suggests this directional diffusion is negligible over the scales of hundred of kilometers considered in this paper. 
\ch{The neglect of wave-wave interactions also fails over large propagation distances. For quadruplet wave interactions, assuming a steepness parameter of $O(0.1)$ and background wave period of $O(10)$ s, the nonlinear interactions are negligible up to a spatial scale at order $O(1000)$ km.}


\ch{Even with non-negligible source terms in the wave action equation, U2H can still provide a useful estimation of the SWH anomalies induced exclusively by  currents. Accurately resolving the corresponding transport terms in the wave action equation introduces additional computational costs, as they call for small time steps and fine angular grids (see figure 7 in \citet{Ardhuin2017}). When such additional computational costs are prohibitive, the U2H framework can provide a fast estimate of the modulations induced by currents.}



W25 note that, because the dependence of the SWH on both the divergent and rotational parts of the current, it is not in general possible to invert the U2H map and infer the current from SWH alone.  It should however be possible to infer some useful statistical properties of the currents. For example, we show in section \ref{sec:aniso} that for \ch{waves with small directional spreading}, the across-wave KE spectra tend to have similar spectral slopes to the along/across-wave rotational KE within their spectral supports. Thus, if the assumptions involved are met, one can infer the slopes of rotational KE spectra from SWH spectra. 



\acknowledgments
This paper is a contribution to the projects M2, T2 and W2 of the Collaborative Research Centre TRR 181 ``Energy Transfers in Atmosphere and Ocean" funded by the Deutsche Forschungsgemeinschaft (DFG, German Research Foundation) - Projektnummer 274762653, which supports HW. 
ABVB is supported by the ONR MURI award N00014-24-1-2554, and NASA awards 80NSSC23K0979 through the International Ocean Vector Winds Science Team and 80NSSC24K1640 through the SWOT Science Team.
JV is supported by the UK Natural  Environment Research Council (grant NE/W002876/1).  WRY  is supported by the National Science Foundation award 2048583. \ch{We thank Jean Bidlot for interesting discussions. Comments from the two anonymous reviewers have improved the manuscript. } 
\datastatement
Codes used for creating synthetic currents, computing the U2H map and plotting are accessible at \url{https://github.com/hannnwang/U2H_examples}.

\appendix[A]
\appendixtitle{Background wavenumber spectrum} \label{app:wavenumberpara} 
The wavenumber dependence described by $f(k)$ in \eqref{bAfD} is as in \cite{WVBYV}. We detail its form to introduce the two parameters that appear, namely the mean frequency $\sigma_m$ and the frequency spread $\delta_\sigma$. We also show that  the U2H map is insensitive to $\delta_\sigma$.  

We specify an SGW energy spectrum of the form $Z(\sigma) D(\theta)$ such that  
\beq \label{gsigmanorm}
\int_0^\infty \int_0^{2\pi} Z(\sigma)D(\theta) \,\dd \sigma \,\dd\theta = g \bHs^2/16.
\eeq
and choose a frequency dependence in the form of the truncated Gaussian
\beq \label{sigspec}
Z(\sigma) = \frac{g\bHs^2\, \ee^{ -(\sigma - \sigma_m)^2/2 \delsig^2}}{8 \sqrt{2 \pi} \delsig \, \erfc(-\sigma_m/\sqrt{2} \delsig)} 
\eeq
with $\sigma > 0$,
which introduces the two parameters $\sigma_m$ and $\delsig$.


The wavenumber dependence $f(k)$ in \eqref{bHs2bA}  is related to $Z(\sigma)$ via $Z(\sigma) \dd \sigma=f(k)\sigma k {\dd k}$ with $\sigma = \sqrt{gk}$. Therefore, 
\beq\label{fk}
f(k)=\frac{g\bHs^2\ee^{ -(\sigma(k) - \sigma_m)^2/2 \delsig^2}}{8\sqrt{2 \pi}\erfc(-\sigma_m/\sqrt{2} \delsig) \delsig \sigma(k) k} 
\frac{\dd \sigma}{\dd k}.
\eeq

The wavenumber dependence $f(k)$ affects the U2H map via the $k$-integral in \eqref{pndef}. Using \eqref{fk}, this integral is computed with help of a symbolic integrator:
\beq
\int_0^{\infty} \!\!\! \!f(k) k^2\,\dd k =\frac{\bHs^2}{16} \sigma_m M\left(\frac{\delsig}{\sigma_m}\right),\,\label{fkk2int1}
\eeq
where
\beq
M(\tau)\defn 1+\sqrt{\frac{2}{\pi}}\frac{\tau\ee^{-\tau^{-2}/2}}{\erfc\left(-1/\sqrt{2}\tau \right)}.
\eeq

The function $M(\tau)$ increases with positive $\tau$. For $0<\tau<0.5$, the value of $M$ ranges in $(1,1.028)$, which corresponds to a variation of less than $2.8$\%, an amount we regard as negligible. For higher $\tau,$ the function $M(\tau)$ increases faster: for $\tau = 1$, $M(1) = 1.29$, meaning that when $\delsig$ is as large as $\sigma_m$, the U2H map's operators can change by $29$\% in  magnitude. 

We assume $\delsig/\sigma_m$ is less than $0.5$ and so neglect the impact of $\delsig$ on the U2H map.  We set $\delsig=0.04$ s$^{-1}$, which  satisfies $\delsig/\sigma_m<0.5$ for the range of $\sigma_m$ considered here. With these values, the truncated Gaussian in \eqref{sigspec} is indistinguishable from a full Gaussian and $\sigma_m$ and $\delsig$ are excellent approximations to the mean and standard deviations.

\appendix[B]
\appendixtitle{Cross-spectrum contribution} \label{app:cross}
We show that provided that the cross-spectrum $\hat C ^{U_{\phi}U_{\psi}}(\bq)$ is isotropic, it plays no role in the isotropic spectrum of the SWH anomaly and hence in the variances of the SWH and of its gradient. Therefore the results of sections \ref{sec:var} and \ref{sec:isospec} hold also for isotropic currents with correlated divergent and rotational components.

In the power spectral relation  \eqref{U2Hspectra}, integration of the cross-spectrum term in \eqref{U2Hspectra} with respect to $\qang$ leads to the term 
\beq \label{coefcross}
2 \Re  \left(\, \hat C ^{U_{\phi}U_{\psi}}(q)/q  \int \hlphi^*(\qang)\hlpsi(\qang) \, \dd \varphi  \right).
\eeq
which we now show vanishes.

The operator $ \hlphi(\qang)$ in \eqref{lparlper1} can be written as
\beq\label{hlpar}
 \hlphi(\qang) =-\frac{32}{g \bHs^2}\bm{P}  \bcdot \eq=-\frac{32}{g \bHs^2} (P_1\cos{\varphi}+P_2\sin{\varphi}).
\eeq
Clearly then, only terms proportional to   $\cos{\varphi}$ and $\sin{\varphi}$ in $\hlpsi^*(\qang)$ can contribute to the integral \eqref{coefcross}. Rearranging expression \eqref{lparlper2} for $\hlpsi(\qang)$ as
\beq\label{hlper}
\hlpsi(\qang)=-\frac{64 }{g \bHs^2} \bm{P} \bcdot \eqp+\frac{16}{g \bHs^2} \sum_{|n|>1} n (-\ii)^{|n|} 2 \pi p_n \, \ee^{\ii n\qang}
\eeq
isolates these term in $\bm{P} \bcdot \eqp$. We then compute
\begin{align}
&\int \hlphi(\qang)\hlpsi^*(\qang)  \, \dd \varphi\\ &=
2\left(\frac{32}{g \bHs^2}\right)^2  \int (\bm{P} \bcdot \eq )(\bm{P}\bcdot \eqp) \, \dd \varphi
\nonumber \\&=2\left(\frac{32}{g \bHs^2}\right)^2  \left(\int P_1P_2 \cos^2{\varphi}  \, \dd \varphi -\int P_1P_2 \sin^2{\varphi} \, \dd \varphi\right) =0
\end{align}
using that $\bm{P}$ is real. 


\appendix[C ]
\appendixtitle{Numerical details of synthetic currents} \label{app:syncur}
For the synthetic currents in section \ref{sec:var},
we draw current samples on square, $2500$ km-wide grids with grid spacing of $2.5$ km. For $2\pi/(200 \text{km}) < q < 2\pi/(6 \text{km})$, the isotropic spectra $\KE_\phi(q)$ and $\KE_\psi(q)$
follow the power laws $q^{\gamma}$. For larger wavenumbers, the spectra are set to decay rapidly. For smaller wavenumbers, they are set as zero in order to alleviate artifacts from a square, non-isotropic spectral grid. The currents are drawn so that the Fourier coefficients of the potential and streamfunction follow zero-mean, random phase distributions, independent at each wavenumber except when the reality condition demands. The potential and streamfunction are independent. 

The synthetic currents shown in figures \ref{fig:realsamplesingleslopes40} and \ref{fig:realsamplesingleslopes1} are drawn similarly. The only difference is that for sake of demonstration, 
we draw a single random function, $\hat \Phi$ say, and take the streamfunction and potential  as $\hat \psi = \sqrt{\alpha} \, \hat \Phi$ and $\hat \phi = \sqrt{1-\alpha} \, \hat \Phi$ so that they are correlated. This choice is made in order that the current speeds remain unchanged in each row, despite different $\alpha$ in figures \ref{fig:realsamplesingleslopes40} and \ref{fig:realsamplesingleslopes1}.

\end{document}